\documentclass{sig-alternate-05-2015}

\usepackage{url}
\usepackage{mathptmx}
\usepackage{amsmath}
\usepackage{times}
\usepackage{xcolor}
\usepackage{lastpage}
\usepackage{amsmath}
\usepackage[utf8]{inputenc}
\usepackage{epsfig,endnotes,color}
\usepackage{array}
\usepackage{multirow}
\usepackage{xspace}
\usepackage{verbatim}
\usepackage{wrapfig}
\usepackage{mathptmx}
\usepackage[english]{babel}
\usepackage{graphicx}
\usepackage{csvsimple}
\usepackage{epstopdf}
\usepackage{subcaption}
\usepackage{enumitem}
\usepackage{datetime}
\usepackage{cleveref}

\usepackage{amssymb}
\usepackage{tabularx}

\newcommand{\projecttitle}{{\scshape PAS-MC}\xspace}

\begin{document}

\title{PAS-MC: Privacy-preserving Analytics Stream for the Mobile Cloud}

\numberofauthors{1}
\author{Josh Joy\qquad Mario Gerla\\UCLA}

\maketitle

\begin{abstract}

In today's digital world, personal data is being continuously collected and analyzed without data owners' consent and choice. As data owners constantly generate data on their personal devices, the tension of storing private data on their own devices yet allowing third party analysts to perform aggregate analytics yields an interesting dilemma.

This paper introduces \projecttitle, the first practical privacy-preserving and anonymity stream analytics system. \projecttitle ensures that each data owner locally privatizes their sensitive data before responding to analysts' queries. \projecttitle also protects against traffic analysis attacks with minimal trust vulnerabilities.We evaluate the scheme over the California Transportation Dataset and show that we can privately and anonymously stream vehicular location updates yet preserve high accuracy.


\end{abstract}

\section{Introduction}
Imagine that you wish to work out at the gym, though you would like to check how crowded the gym is before hand. Should everyone currently at the gym be required to publicly disclose their sensitive location information every minute in order to allow us to check how crowded a particular gym is? Clearly this is a privacy concern. The question we seek to answer is how to harvest sensitive data (e.g., current location) from mobiles in real-time while simultaneously providing strong privacy guarantees.

Recently researchers have taken a renewed interest in data privacy. The Netflix Prize privacy fiasco is a recent example where public disclosure of a large data set was not properly sanitized. This privacy oversight resulted in Netflix being sued~\cite{netflix-privacy-lawsuit} as several individuals being de-anonymized. The literature describes a number of mechanisms for protecting user privacy while allowing useful data analytics~\cite{Sweene02,LiLV07,MachanavajjhalaGKV06,DBLP:conf/tcc/DworkMNS06}. Among these, differential privacy has gained broad acceptance. It adds differential private noise to the aggregate query results in such a way as to hide the presence or absence of any individual user. A key strength of this approach is that it provably protects the privacy of individual users.


Currently, we can think of three different data harvesting models whereby sensitive data is collected for analytics. The first is a centralized model whereby companies today collect sensitive location information into centralized data repositories. For various reasons (e.g., regulatory mandates, privacy policies or ethic policies) this sensitive data is only collected and not publicly disclosed. There are no strong privacy guarantees. Confidence in the protection of sensitive data is left to regulatory policies. Additionally, these centralized data repositories are central points of trust vulnerabilities which continue to invite data breaches \cite{violation1,violation2,violation3,violation4,violation5,violation6}. 

The second is the original differential privacy mechanism that assumed a trusted central database maintaining all users' personal data~\cite{DBLP:conf/icalp/Dwork06,DBLP:conf/tcc/DworkMNS06}. Data is then privatized upon disclosure (either upon release from the centralized repository or by the third party receiving the data). An example of this is offered by Google Waze which collects real-time vehicle locations and has a data sharing agreement with multiple city agencies ~\cite{googlewazeconnectedcitizens}. However, data owners must make strong trust assumptions with a centralized database. Namely, the database must be trusted to halt all queries once the differential privacy budget has been exhausted. Also, data owners do not have granularity in consent and control mechanisms over the release of their personal data. In the past, trust in centralized systems safeguarding privacy has been violated~\cite{breach1,breach2}.


Finally, the last is a distributed model, used in \projecttitle, which enhances the prior two models and ensures that data owners release a privatized and anonymized version of their sensitive location information. Data owners maintain complete control and consent over the utilization of their sensitive data. Upon receiving a query from a remote  analyst, the data owner queries its local database and produces a truthful answer. In this model, data collection agencies maintain and store only privatized data thus mitigating any potential trust vulnerabilities.  Privacy-preserving distributed stream monitoring systems have been proposed before~\cite{DBLP:conf/ndss/FriedmanSKS14,DBLP:conf/pet/ChanLSX12,DBLP:conf/eurocrypt/DworkKMMN06}.  However, they all require some form of synchronization, or are tailored for heavy-hitter monitoring only (i.e., they can only report on a fraction of the data).

This paper presents \projecttitle the first practical, anonymity and privacy-preserving stream analytics system that collects data in real-time from mobiles and provides strong privacy guarantees. In \projecttitle, each data owner's personal data resides on the data owner's own device.  Once receiving a query, each data owner does not directly respond to the query with the truthful answer.  Instead, the data owner locally privatizes their answer based on the randomized response mechanism~\cite{warner1965randomized,fox1986randomized} such that only privatized data is released (rather than the original answer).  Randomized response satisfies the local differential privacy requirement such that each data owner's response is independently differentially private, regardless of the amount of differential privacy noise added by other data owners or system components. That is, for a response of "Yes" the data owner has an equivalent probability of having or not having the sensitive attribute \S\ref{sec:randomizedresponse}. Thus, randomized response eliminates the need for strong trust assumptions regarding the aggregation mechanism in a distributed setting. Additionally, there is no need to synchronize data owners (e.g., no MPC amongst data owners) or other system components for adding the appropriate differential private noise, leading to low latency and achieving real-time analytics.

To anonymously transmit the data owners' randomized responses to a data aggregator, each data owner generates functional secret shares (FSS) \cite{DBLP:conf/eurocrypt/BoyleGI15}. FSS slices the response into multiple shares. Then, each share is individually transmitted to an independent aggregator. Each aggregator independently and asynchronously processes each share. At the end of an agreed upon epoch, all aggregators share their results. As long as there is at least one honest aggregator, the data owners' anonymity is guaranteed.  Eventually, the aggregators generates the query result based on the received randomized and anonymized responses, and transmits the query result to the appropriate analyst. Thus, it is not possible for a malicious adversary to discover which data owner transmitted a particular randomized response from within the anonymity set. FSS hardness assumptions does not depend on a particular pseudorandom number generator (as opposed to a homomorphic pseudorandom generator \cite{DBLP:conf/sp/Corrigan-GibbsB15}) which allows \projecttitle to be efficient and scalable seen as shown in \S\ref{sec:evaluation:scaling}. We can privately and anonymously stream 220,000 data owners' location data with minimal probability of collision with a key size of 112KB.
 
To evaluate our privacy-preserving approach we examine a vehicular case study utilizing the California Transportation Dataset. We examine both rush hour and off peak to demonstrate our privacy mechanism works well for small sample sizes \S\ref{sec:evaluation:accuracy}.

In this paper, our contribution is a software which that the first time achieves all of the following for a real-time system:

\begin{enumerate}
\item a privacy scheme that allows each data owner to operate independently without coordination or a centralized service, 
\item a scalable anonymity system resistant to traffic analysis, and
\item disruption protection for the anonymity scheme
\end{enumerate}

\begin{figure*}[ht!]
\subcaptionbox{Privatization\label{fig:rushhour}}{\includegraphics[width=1\columnwidth]{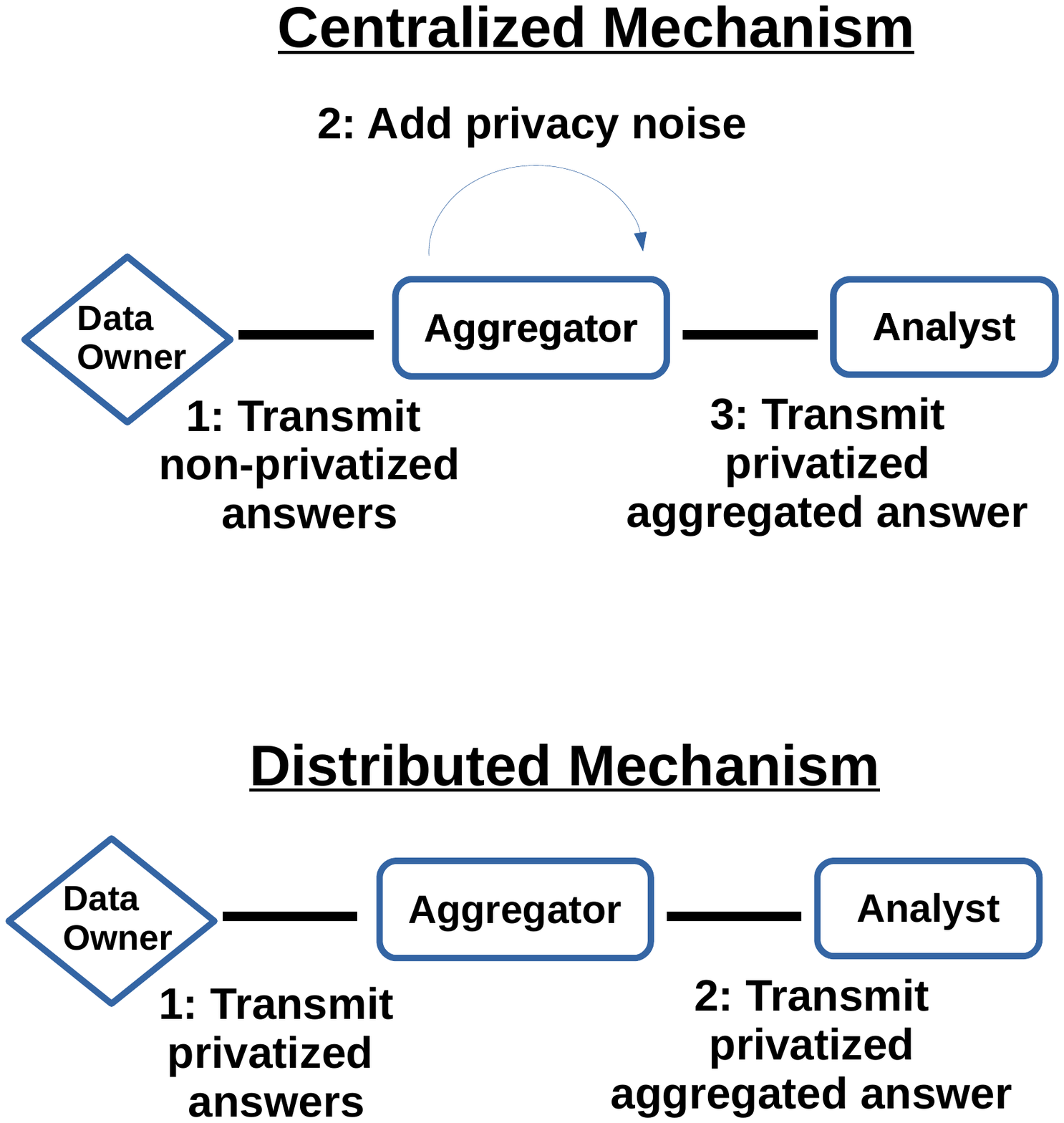}}\hfill
\subcaptionbox{System Flow\label{fig:offpeak}}{\includegraphics[width=1\columnwidth]{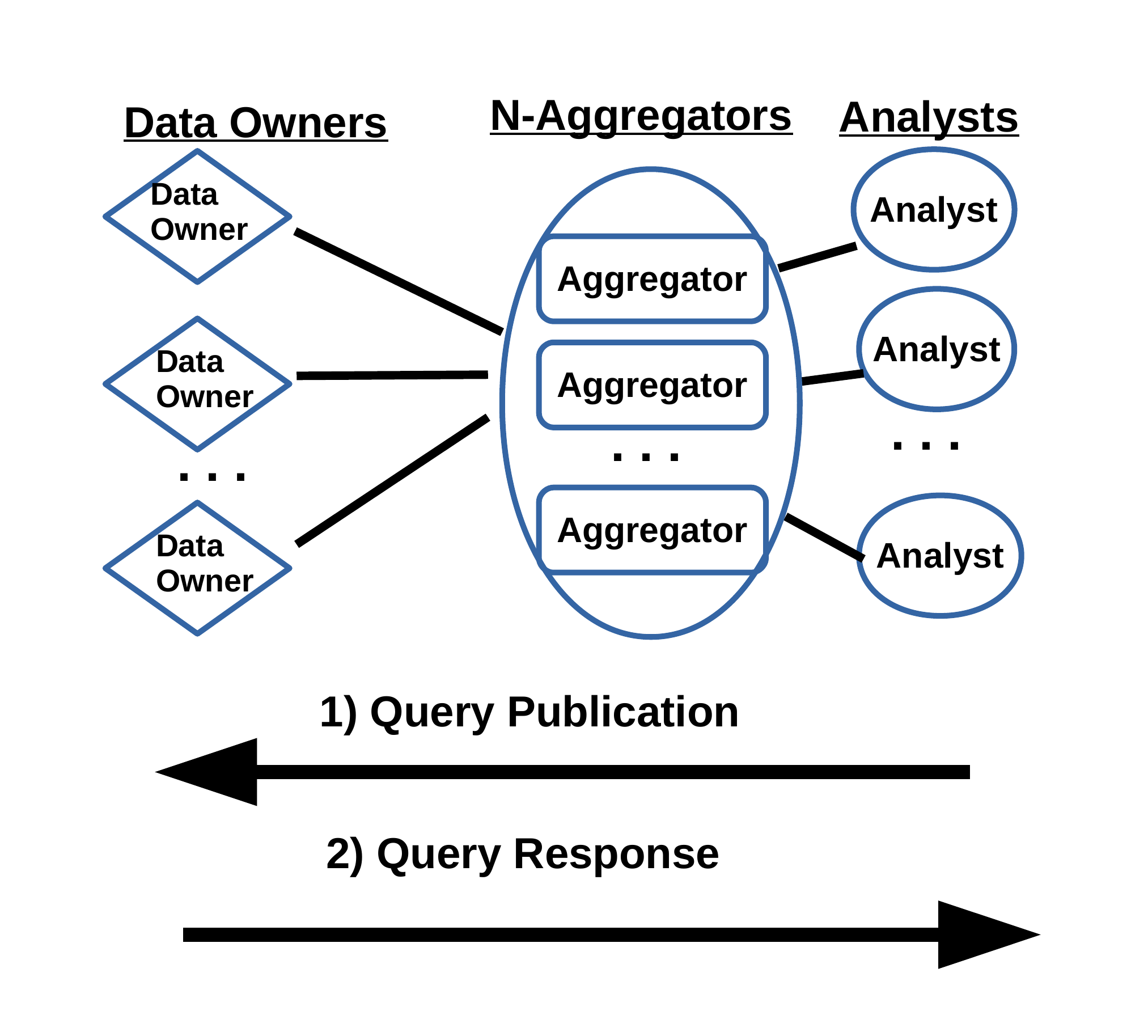}}\hfill
\protect\caption{\projecttitle system overview. (a) illustrates that the data owner has control and consent over the privatization release as opposed to the centralized mechanism which requires strong trust assumptions regarding the aggregator adding differential private noise.}
\label{fig:architecture}
\end{figure*}

\section{Goals and Problem Statement}

We now describe the system goals, performance goals, threat model, and privacy goals of \projecttitle. Figure~\ref{fig:architecture} shows an overview of the flow of queries and responses and demonstrates that the privatization occurs before reaching the aggregator.

\subsection{System Goals}

The system should support analysts who wish to run a population study. The analysts issue a query for those interested data owners that privately and anonymously reply. Analysts are able to formulate long-standing signed queries. These queries continually elicit privatized responses during the defined query epoch. The analysts are deemed to be reputable, e.g., Department of Transportation, National Institutes of Health, or Centers for Disease Control. Each analyst controls an aggregation server.

We use a vehicular example as motivation. Vehicles today have an array of sensors that collect position, speed, acceleration, and CO2 values. Analysts (e.g., researchers, municipalities, automakers, insurance companies) are able to utilize this information to study road and traffic patterns for autonomous vehicles. For example, real-time traffic information and CO2 levels allow municipalities to reroute autonomous traffic to reduce traffic congestion and to reroute heavy polluters to reduce pollution levels. While real-time mobile crowdsourcing has high utility and benefit to society, real-time sensor data harvesting has serious privacy risks. Autonomous passengers (data owners) do not wish to be constantly tracked, as this violates location privacy.

The queries are propagated using roadside WiFi units or LTE repeaters. The requests may also be piggybacked on responses to traffic information requested by the driver, or in response to periodic driver polls. To reduce traffic O/H, or in response to periodic driver polls the queries may be posted to an edge website that mobiles of a certain class frequently check (e.g., taxicab website).

The long-standing queries are needed to be fetched only once by each data owner. Data owners have the choice of answering the query and can configure their devices to respond if certain conditions are met (e.g., whitelisting or blacklisting certain sensors or response frequency).

The responses of data owners and aggregation processing proceeds in epochs. That is every epoch each data owner privately and anonymously transmits their respective answer to the aggregator servers. The aggregation servers then compute the final aggregate using the received responses within this epoch. Epochs are defined on the order of seconds.

As there is a tradeoff between privacy and utility, the system should strike a superior balance the tradeoff of strong privacy and utility. We strive for a small percentage of error for both sparse and large datasets.

\subsection{Performance Goals}

The system should scale gracefully as compared to related anonymity systems which provide strong guarantees against traffic analysis. Network bandwidth should be minimal as compared to existing bandwidth hungry systems \cite{DBLP:journals/joc/Chaum88,DBLP:conf/sosp/HooffLZZ15}.

\subsection{Threat Model}

Aggregation servers may try to collude, though we assume there is at least one honest aggregation server. Each aggregation server is owned by a set of distinct reputable analysts.

Aggregation servers are expected to be available and online, so we do not consider denial of service attacks whereby data owners are not able to transmit their responses. We assume aggregation servers are honest-but-curious, i.e., servers do not corrupt the messages though can attempt to read all messages.

In the mobile environment, users frequently go online and offline. Thus, a malicious observer can perform traffic analysis techniques by observing the sender, receiver, and frequency of messages. The malicious observer essentially performs performs "intersections" across the sets of active users in order to de-anonymize them \cite{DBLP:conf/pet/MathewsonD04,DBLP:conf/ih/DanezisS04}. Intersection attacks across long durations (many epochs) are very difficult to overcome for anonymity systems. We do not consider these attacks in this paper. However, \projecttitle works within a single epoch and our scheme scales to the order of hundreds of thousands of data owners; thus, making it increasingly difficult to execute these intersection attacks.

There are two pollution attacks we consider: a malicious data owner who repeatedly answers a query in order to inflate the aggregate sum and a malicious data owner who repeatedly answers a query within a single epoch. To prevent a single answer, such as a large number, from distorting the aggregate sum, we utilize a bit vector response which limits the data owner to only replying '0' or '1'. To prevent multiple responses within a single epoch, data owners are limited to a single response per epoch per query. Thus, malicious data owners which try to provide spurious answers will remain limited in their impact and can be eventually detected and blacklisted (certification revoked).

We do not consider malicious analyst who provide falsified aggregated results. Economic incentives may help defer this. We also look to address this in future work.

\subsection{Privacy Goals}

We assume all queries are signed and from reputable analysts. This provides provenance in the case of a dishonest analyst that may formulate a specially crafted query that attempts to deprivatize a data owner. Analysts may publicly publish the aggregated answers as the answers are differentially private.  Malicious analysts do not gain any advantage from creating multiple sybil analysts, as the data owners privacy deficit will be decremented for each answer and data owners are able to detect duplicate queries.

Data owners' privatized location responses should leak no more data than if they were not participating in the population study. Each data owner retains their own data on devices that they control and manage. The data owners then choose to participate in responding to each query. All responses before they leave the data owner are privatized and anonymized. The anonymization mechanism requires only a single honest aggregation server to participate and that there are at least two honest data owners. The privacy mechanism should satisfy the local differential privacy criteria. Thus, there is no centralized or trusted aggregation mechanism that adds differentially private noise. Moreover, neither servers nor data owners can collude to deprivatize the data. 

Our goal for anonymity is that a data owner is able to transmit a message such that the message is unable to be linked back to the data owner. That is, a data owner is anonymous within a group of data owners, i.e., the anonymity set. The anonymity scheme should also be robust to traffic analysis. We rely on a public-key infrastructure (PKI) to thwart sybil attacks. However, the use of PKI does not preclude anonymity, as data owners remain anonymous within the anonymity set.

The larger the anonymity set, the stronger the anonymity guarantees are for the data owner. For example, if the anonymity set size is only a dozen, it is straightfoward to either coerce or reprimand the small group. The idea is that there is strength in numbers, and our goal is to build anonymity sets on the  order of hundreds of thousands. Large scale anonymity sets are more difficult to coerce and breach.

\section{Preliminaries}

\subsection{Differential Privacy}

Differential privacy has become the \emph{gold standard} privacy mechanism which ensures that the output of a sanitization mechanism does not violate the privacy of any individual inputs.  A privacy mechanism $San()$ provides $\epsilon$-differential privacy~\cite{DBLP:conf/icalp/Dwork06,DBLP:conf/tcc/DworkMNS06} if, for all datasets $D_1$ and $D_2$ differing on at most one record, and for all outputs $O \subseteq Range(San())$:
\begin{equation}
\Pr[San(D_1) \in O] \leq e^{\epsilon} \times \Pr[San(D_2) \in O]
\label{eqn:dp}
\end{equation}

That is, the probability that a privacy mechansim $San$ produces a given output is almost independent of the presence or absence of any individual record in the dataset.  In other words, it is difficult to determine whether any individual record is in the dataset, thus protecting privacy.  The privacy parameter $\epsilon$ controls the tradeoff between the accuracy of a privacy mechanism and the strength of its privacy guarantees: smaller $\epsilon$ provides stronger privacy but lower accuracy, and vice versa.

\subsection{Obfuscation}
\label{sec:obfuscation}

A point function, without loss of generality, for a given input $x$ outputs $y$. That is, $f(x)=y$ for the special input $x$ and $f(\neg x)=0$ for all other inputs.

Point functions are very powerful as it can be shown certain classes of cryptographic obfuscation can be achieved. That is, given an obfuscator $O$ for a program $P$, an obfuscated new program $P'$ will be produced which has the same functionality as $P$ though has some definable notions of "opaqueness". This has numerous cryptographic applications from software protection to homomorphic encryption \cite{DBLP:journals/jacm/BarakGIRSVY12}.

In our case, we are interested in a variant of the point function called a distributed point function. A distributed point function is a keyed family function that generates multiple shares such that evaluating the combination of all the shares with the special input $x$ yields the output $y$, and 0 for all other inputs. That is, given $n$ keyed family shares $f(x_{k_1}) \oplus f(x_{k_2}) ... \oplus f(x_{k_n}) = y$. 

Distributed point functions (DPF) also have many applications. DPFs are useful for private information retrieval and private information storage \cite{DBLP:conf/eurocrypt/GilboaI14,DBLP:conf/sp/Corrigan-GibbsB15}. We utilize distributed point functions to achieve anonymous writes such that a data owner can anonymously $write$ a single $value$ at a single $address$, that is perform $write(address,value)$.

\section{System Architecture}

The system, as seen in Figure~\ref{fig:architecture}, is composed of data owners who locally privatize and anonymize their answers before transmitting to the aggregators in response to queries by analysts. We now describe the data owner privatization mechansims which utilize the randomized response mechanism and then the anonymization mechanism which utilizes a technique called function secret sharing.

\subsection{Privacy Mechanism}
\label{sec:randomizedresponse}

To ensure that each data owner individually and indepedently privatizes their answer, we utilize the randomized response mechanism. Randomized response~\cite{warner1965randomized} was originally created by social scientists as a mechanism to perform a population study over sensitive attributes (such as drug use or certain ethical behaviors). Randomized response allows data owners to locally randomize their truthful answer to analyts' sensitive queries and respond only with the privatized (locally randomized) answer. We utilize randomized response as our privacy mechanism as randomized response satisfies the differential privacy guarantee for individual data owners (see \S\ref{sec:priv_g}) and  it provides the optimal sample complexity for local differential privacy mechanisms~\cite{DBLP:conf/nips/DuchiWJ13}.

So far, the randomized response mechanism has been widely adopted by both social scientists and computer scientists \cite{warner1965randomized,DBLP:conf/ccs/ErlingssonPK14}. There are many different randomized response mechanisms in the literature.  In this section, we present only the mechanism described in~\cite{fox1986randomized} because it strikes a superior balance between the utility and the privacy guarantee of randomized responses, as compared to other mechanisms~\cite{warner1965randomized,kuk1990asking,greenberg1969unrelated,greenberg1971application}.

\subsubsection{Mechanism Description} We will now describe how each data owner privatizes their response utilizing the randomized response mechanism. Suppose each data owner has two independently biased coins. Let the first coin flip heads with probability $p$, and the second coin flip heads with probability $q$.  Without loss of generality, in this paper, heads is represented as ``yes'' (i.e., 1), and tails is represented as ``no'' (i.e., 0).

Each data owner flips the first coin. If it comes up heads, the data owner responds truthfully; otherwise, the data owner flips the second coin and reports the result of this second coin flip.  

Suppose there are $N$ data owners participating in the population study. Let $\hat{Y}$ represent the total aggregate of ``yes`` randomized answers. The estimated population with the sensitive attribute $Y_A$ can be computed as:
\begin{equation}
\label{eqn:yo}
Y_A = \frac{\hat{Y} - (1 - p) \times q \times N}{p}
\end{equation}

The intuition behind randomized response is that it provides ``plausible deniability'', i.e., any truthful answer can produce a response either ``yes'' or ``no'', and data owners retain strong deniability for any answers they respond. If the first coin always comes up heads, there is high utility yet no privacy. Conversely, if the first coin is always tails, there is low utility though strong privacy. As we will show in the evaluation (see \S\ref{sec:evaluation}), by carefully controlling the bias of the two coin flips, one can strike a balance between utility and privacy.

\subsubsection{Multiple Sensitive Attributes}

While randomized response is an intuitive privacy mechanism for a single sensitive attribute, naturally the question becomes how does one deal with multiple sensitive attributes?  A host of "polychotomous" mechanisms have been studied and surveyed in the literature \cite{fox1986randomized} using multiple randomizing mechanisms or maximum likelihood estimators \cite{doi:10.1080/01621459.1981.10477741}. However, it turns out that simply repeating an application of \cite{fox1986randomized} for each sensitive attribute turns out to be an ``optimal'' \cite{doi:10.1080/01621459.1981.10477741} approach.

Thus, \projecttitle repeats the randomized response mechanism for each sensitive attribute. For example, if a traffic analyst wishes to understand the traffic flow of a few key locations, the traffic analyst issues a query that is a Boolean bit-vector asking each data owner to indicate the location they are at. Then, each data owner performs randomized response for each location and replies with a Boolean bit-vector. The traffic analyst then aggregates and sums the bit-vectors to calculate the number of vehicles at each location.

\subsubsection{Utility of Randomized Response}

We use two metrics to evaluate the utility, root mean squared error (RMSE) and relative error. The RMSE gives us an idea of how much of a fluctuation the privacy mechanism provides over the population. Relative error gives us an idea of how accurate the mechanism can be on average.

Suppose the estimated and actual counts of the population of data owners with the sensitive attribute are $Y$ and $Y_A$, respectively.  The utility defined as the RMSE is the square root of the mean of the square of the errors.
\begin{equation}
\label{eqn:rmse}
RMSE(Y) = \sqrt( \frac{1}{N} \sum_{i=1}^{N} (Y_A - Y)^2 )
\end{equation}

Here, a smaller RMSE means that the aggregate privatized responses will be less likely to vary from the aggregate actual responses.

Then, the utility defined as the relative error $\eta$ --- the magnitude of the difference between the actual count and the estimated count, divided by the magnitude of the actual count.
\begin{equation}
\label{eqn:eta}
\eta = \bigl|\frac{Y - Y_A }{Y}\bigr|
\end{equation}

Here, smaller relative error $\eta$ means higher utility of the randomized responses, and vice versa.

\subsection{Privacy Guarantee of Randomized Response}
\label{sec:priv_g}

Our privacy goal is for a data owner's response to minimize information leakage. That is, a ``yes'' response $Y$ by a data owner should be as equally likely if the data owner does or does not have the sensitive attribute $A$.
\begin{equation}
\label{eqn:conditional}
\Pr[A \mid Y] = \Pr[\neg A \mid Y] 
\end{equation}

It turns out that by carefully controlling each coin bias we can control this privacy parameter. In our evaluation results we show for particular values of $p$ and $q$, the probabilities of having or not having the sensitive attribute are equally likely \S\ref{sec:evaluation:leakage}.

Additionally, based on expression~\ref{eqn:dp}, the randomized response mechanism can achieve $\epsilon$-differential privacy, where:
\begin{equation}
\epsilon = \ln(\frac{\Pr[\textrm{Y \textbar A}]} {\Pr[\textrm{Y \textbar $\neg$A}]})
\end{equation}

\noindent Or,
\begin{equation}
\epsilon = \ln(\frac{\Pr[\textrm{Y \textbar $\neg$A}]} {\Pr[\textrm{Y \textbar A}]})
\end{equation}

\noindent Whichever is larger.
More specifically, the mechanism~\cite{fox1986randomized} achieves $\epsilon$-differential privacy, where:
\begin{equation}
\label{eqn:e-forced}
\epsilon = \ln ( \frac{p + (1 - p) \times q}{(1 - p) \times q} )
\end{equation}

That is, if a truthful answer is ``yes'', then the randomized answer will be ``yes'' with the probability of `$p + (1 - p) \times q$'.  Else, if a truthful answer is ``no'', then the randomized answer will become ``yes'' with the probability of `$(1 - p) \times q$'.

\subsection{Discussion of Privacy Approach}

It is important to note the following, the queries are signed and are publicly posted. This ensures that the analysts are held accountable and are easily auditable. Additionally, data owners have the ability to view and inspect the query before agreeing to participate and response to the query. It is easy to visualize location queries by displaying a map with the points of interests clearly marked. Since the points of interests should be very busy areas, there is little concern that the data owner would be the only individual at a particular point reporting their location. Additionally, since an analyst single query is over multiple locations, the data owners' response using a boolean bit vector, the data owner may potentially report themselves at multiple locations due to the local differential privacy mechanism. However, in the aggregate the noise is minimal as we show in the results (see \S\ref{sec:evaluation:accuracy}).

\subsection{Anonymous Data Upload}

We utilize function secret sharing (FSS) \cite{DBLP:conf/eurocrypt/BoyleGI15} to anonymously upload a message with traffic analysis protection. FSS cryptographic properties hold as long as there is at least one honest server which does not collude.

The intuition behind FSS to achieve anonymity is as follows. Recall the distributed point function (DPF) where only one specific input has an output value and all other inputs are zero \S\ref{sec:obfuscation}. Using the DPF, each key is sent to a separate server such that a single key or even $n-1$ keys does not leak the corresponding $(x,y)$ pairing. Each server then separately evaluates its key over all possible inputs. These "evaluations" are then combined with the evaluations of the other servers to finally generate the $(x,y)$ pair. Naturally, if there is only one user it's straightforward to discover what $(x,y)$ pair was used. However, as long as there are two or more users, then it's not possible with chance better than random to discover which user corresponds to which $(x,y)$ pairing.

We now explain FSS in further detail. Suppose we wish to secretly share a function with \textit{p} parties where at least one party is honest. Suppose there is an input \textit{x} which is \textit{n} bits and the output \textit{y} which is \textit{m} bits. Given \textit{p} keys such that the strings are randomly sampled from the space of ${\{0,1\}}^{2^n*m}$ (total number of inputs multiplied by the size of the message), these strings should evaluate to the message \textit{m} whereby $f(x)=y$ such that $\bigoplus_{i=j}^p k_i[x]=y$. Thus, in this case $p-1$ parties are unable to XOR their keys to discover $f(x)$.

As long as two or more users do not choose the same input $x$, each user is able to write their respective message $y$ to input $x$. Each user proceeds by sending their keys to each respective server. Each server then performs a bitwise XOR of the evaluation of every $f(x_k)$ over the received key $k$ such that $\bigoplus_{x=0}^{2^{n}-1} f(x_k)$. That is there is a total of $2^n$ evaluations at each server for each key. Each result of $f(x_k)$ is XORed locally at each server resulting in an intermediate computation. This intermediate computation is then shared with each other at the end of the agreed epoch.  \linebreak[3] \newline

\textbf{Intermediate Results}:

\begin{equation}
Intermediate~Results~Server_1= \bigoplus_{x=0}^{2^{n}-1} f(x_{k_1}^a) \oplus f(x_{k_1}^b) ... \oplus f(x_{k_1}^p)
\end{equation}

\begin{equation}
Intermediate~Results~Server_2= \bigoplus_{x=0}^{2^{n}-1} f(x_{k_2}^a) \oplus f(x_{k_2}^b) ... \oplus f(x_{k_2}^p)
\end{equation}

\textbf{Final Output}:

\begin{equation}
Output~Server_1=Intermediate~Results~Server_1 \oplus Intermediate~Results~Server_2
\end{equation}

\begin{equation}
Output~Server_2=Intermediate~Results~Server_2 \oplus Intermediate~Results~Server_1
\end{equation}

\textbf{Servers' Output Should Match}
\begin{equation}
Output~Server_1=Output~Server_2
\end{equation}

Additional details regarding the cryptographic techniques and proofs can be found in the FSS paper \cite{DBLP:conf/eurocrypt/BoyleGI15}.

\subsection{Disruption Protection}

To protect against malformed FSS shares whereby a malicious data owner may attempt to define an output at multiple input values, we utilize multiparty computation (MPC). The FSS shares uploaded to the servers are then verified by the below MPC protocol between the servers.

MPC allows a public function to be computed by multiple parties using private inputs, such that each party only knows its own input and the output of the function, nothing else is revealed \cite{DBLP:conf/stoc/GoldreichMW87}. By having the servers perform MPC amongst themselves, we can uphold the data owner anonymity guarantees as well verify that the shares are properly formed without relying on data owner coordination and synchronization. Invalid FSS shares can be quickly XORed out of the intermediate results once they are found, thus eliminating the pollution from the results.

A valid set of FSS shares would be those that only for the special input $x$ does $f(x)=y$ and for all other values of $x$ equals $0$.  The following MPC protocol is robust up to  $n-1$ players being corrupted, so we assume a computationally bounded adversary and rely on public-key cryptography.

A straightforward manner to verify the FSS shares is for each server to evaluate the entire input space over a single FSS share and then XOR the results together and ensure only one input has a non-zero output and the remaining outputs are zero. However, this clearly breaks the anonymity property as all servers know both 1) which data owner sent a corresponding share and 2) what the output of that particular share is. MPC ensures that the only output revealed from evaluating the shares is whether it satisfies the point function rather than revealing the data owner's message.

We now describe the MPC protocol as follows:

\begin{enumerate}
\item Each server evaluates its given FSS share over the total input space
\item For each evaluation result, each server generates a random key and XORs this value with the evaluation result generating an encrypted value
\item Each server then performs a MPC XOR with these values
\item Each server then performs a NXOR over the result and checks if the value is equal to $0$. If its equal to $0$, the result is $1$, else the result is $0$
\item The results of all the NXORS are then summed
\item The final output of the function is whether the sum of the results equals to $n-1$, that is there should be $n-1$ $0's$
\end{enumerate}

The nice part about this computation is that the public function needs mainly XOR operations and not costly AND operations. Also, pre-processing and other recent MPC improvements can be used to ensure efficient computation \cite{DBLP:conf/crypto/LindellPSY15,DBLP:conf/esorics/DamgardKLPSS13}.
\section{Evaluation}
\label{sec:evaluation}

\subsection{Privacy Leakage}
\label{sec:evaluation:leakage}

To understand the information leakage that may occur when a data owner responds ``Yes'', we evaluate the conditional probability. We wish to understand when a data owner answers ``Yes'', what is the probability that the data owner may have or may not have the sensitive attribute. Adjusting the coin flip bias must be done with consideration of the estimated population fraction that actually does have the sensitive attribute. For example, if a large majority of the population has the sensitive attribute, adjusting the first coin flip $p$ to a large number such as $0.9$ will leak a large deal of information as the majority of data owners will respond truthfully. However, as it turns out, if only a small minority of the population has the sensitive attribute in question, then the first coin flip $p$ can be as large as $0.9$.

For our evaluation, there are a total of 222,704 vehicles with a maximum of 860 vehicles and a minimum of 1 vehicle at a station. With a total of 1,157 stations at rush hour, we take the underlying fraction of vehicles at a given station to be 0.005. We evaluate the conditional probability of whether or not a data owner has or doesn't have a given sensitive attribute using the coin flip biases of  $p$=0.995 and $q$=0.999. The values are given in Table~\ref{tab:leakage} and show that the two conditional probabilities are essentially equivalent. That is, the information leaked is negligible.

\begin{table*}[]
\centering
\begin{tabular}{|p{1cm}|p{1.0cm}|}
\hline
\textbf{P(A$\mid$Y)} & 0.501502                   \\ \hline
\textbf{P($\neg$A$\mid$Y)}  & 0.498498                 \\ \hline
\textbf{$\epsilon$} & 5.299313				\\ \hline
\end{tabular}
\caption{Let Y be the privatized yes response. Let A be whether the data owner has sensitive attribute (data owner is at a particular station). These values are for $p$=0.995 and $q$=0.999.}
\label{tab:leakage}
\end{table*}

\subsection{Accuracy of Privacy Mechanism}
\label{sec:evaluation:accuracy}

To understand the accuracy of our privacy mechanism of randomized response, we evaluate the proposed scheme over the California Transportation Dataset\cite{pems}. The particular dataset we utilize collects traffic flow count from under the surface loop detectors in Sacramento freeways \cite{cwwpinformation}.

There are about one thousand stations in this particular district, and 222,704 vehicles total. We examine both peak (5pm) and off-peak (3am) traffic times to understand the impact of sparse and large datasets on our algorithm. We use RMSE and relative error to evaluate the utility of our approach. The resulting metrics error are calculated by taking the average error values over each station. 

The reporting mechanism works as follows. The query is formulated over every station, i.e., the query is a Boolean bit vector with each bit represented by an individual station. Each vehicle knows its location. The vehicle's actual answer should be True only at a single index. However, each index in the bit vector represents a sensitive question asking if a vehicle is at a particular station. Each vehicle then performs randomized response over each station (represented by in index in the bit-vector).

\begin{figure*}[!tb]
\subcaptionbox{Rush Hour\label{fig:rushhour}}{\includegraphics[width=1\columnwidth]{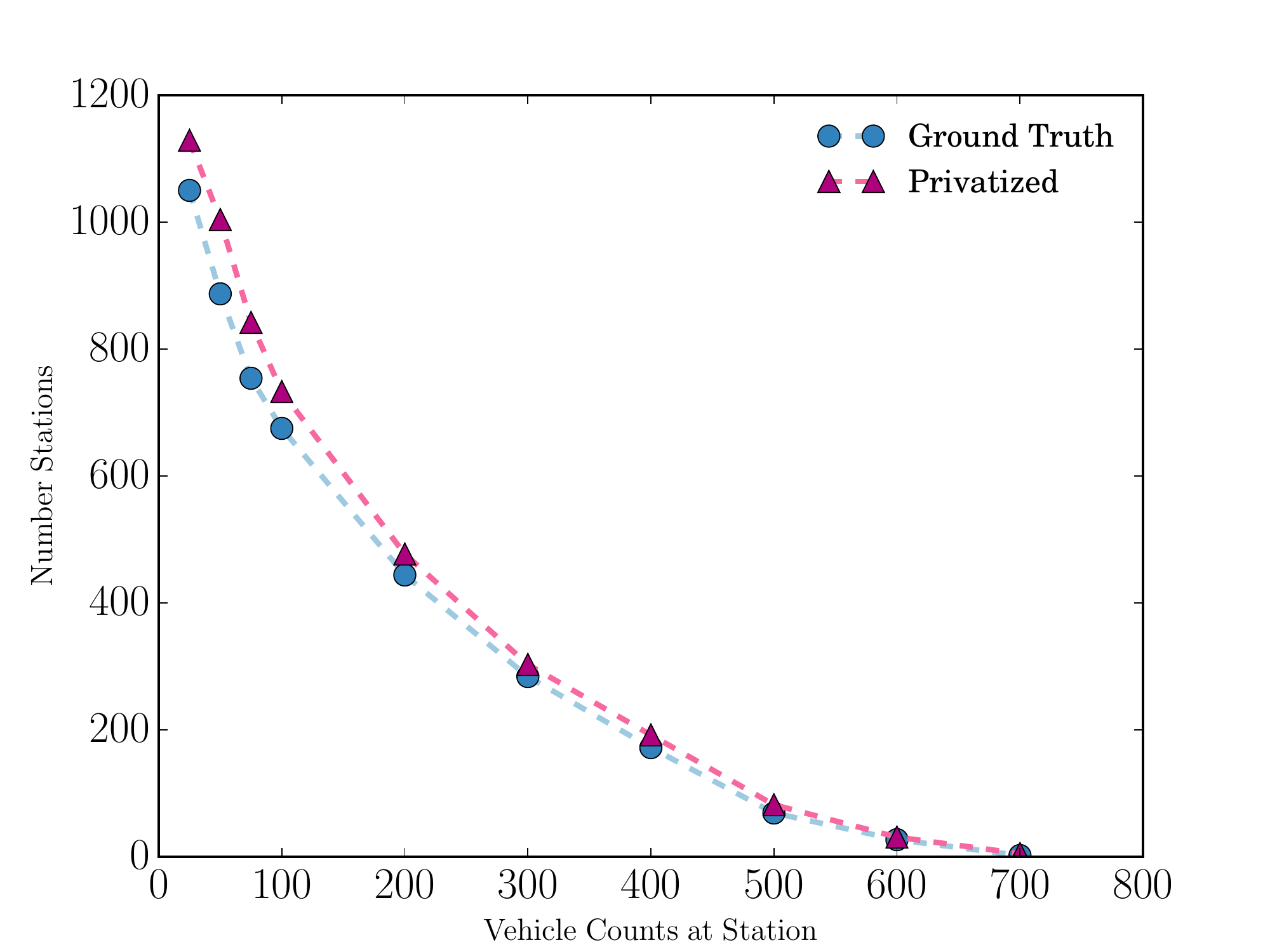}}\hfill
\subcaptionbox{Off Peak\label{fig:offpeak}}{\includegraphics[width=1\columnwidth]{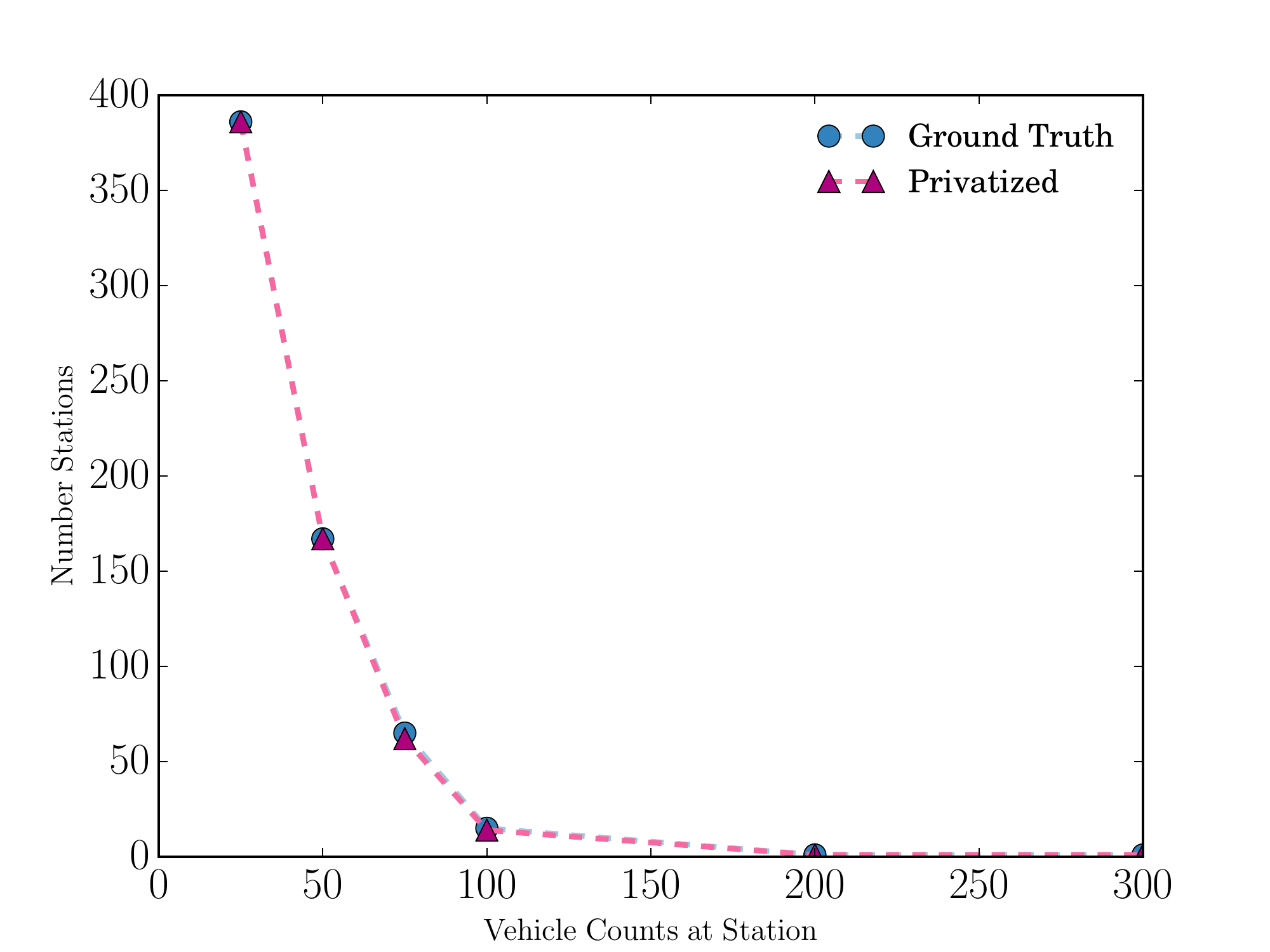}}\hfill
\protect\caption{Randomized response evaluation for vehicular counts for each vehicle tracking station.}
\label{fig:caltrans}
\end{figure*}

Figure~\ref{fig:caltrans} shows the results. Table~\ref{tab:caltran} shows that the off-peak relative error matches the performance of the peak traffic and that the RMSE is small during off-peak hours. This has several implications. The first is that we are more interested in performing traffic rerouting during congestion periods and we can reasonably expect that we can perform this with low relative error. The second is that we can achieve our goal of large anonymity sets on the orders of hundreds of thousands.

\begin{table*}[]
\centering
\begin{tabular}{|p{1cm}|p{1.0cm}|p{2cm}|p{2cm}|}
\hline
                         & \textbf{\# Stations} & \textbf{Avg \newline Relative \newline Error} & \textbf{Avg RMSE} \\ \hline
\textbf{Rush hour (5pm)} & 1157                  & -0.068742           &  25.091603 \\ \hline
\textbf{Off hour (3am)}  & 1017                  & -0.067416           &  5.288476 \\ \hline
\end{tabular}
\caption{Randomized response evaluation for vehicular counts for each vehicle tracking station.}
\label{tab:caltran}
\end{table*}

\subsection{Scaling Effects of Anonymous Data Upload}
\label{sec:evaluation:scaling}

We have implemented \projecttitle using the golang programming language. The server is running Ubuntu 14.04.1 with 6-core Intel Xeon E5-2420 @ 2.20GHz CPUs. We use AES-based PRNG.

To understand the scaling effects, we compare with Riposte's $p$-party protocol which also utilizes a distributed point function. Table~\ref{tab:dpf} shows how \projecttitle is able to quickly scale using the FSS primitive. Riposte is limited by the DDH-based pseudorandom generator (PRG), while \projecttitle is not restricted to a particular PRG thus allowing for much higher throughput. Additionally, the algorithm is extremely parallelizable so adding additional servers will see a corresponding throughput increase.

\begin{table*}[]
\centering
\begin{tabular}{|p{2.3cm}|p{6cm}|p{6cm}|}
\hline
						& \textbf{8 party}		& \textbf{10 party}		\\ \hline
\textbf{Riposte} 			& 1 write every 3.44 seconds (8 server cluster, 1,024 row DB)	& 3 writes every second (10 server cluster, 64 row DB)             	\\ \hline
\textbf{\projecttitle}  		& 9 writes every second (1 server cluster, 1,024 row DB)	& 42 writes every second (1 server cluster, 512 row DB)           		\\ \hline
\end{tabular}
\caption{The message size is 160 bytes. We evaluate \projecttitle on a single server. }
\label{tab:dpf}
\end{table*}

\subsubsection{Deployment:} 

Using the $10$-party protocol, we can handle a database size of $512$ with $10$ servers in about $1$ second. To scale to $220,000$ users would require about $430$ clusters. A total of $4,300$ servers can handle $220,000$ users in one second. A total of $70$ servers can handle $220,000$ users in one minute.

As the FSS primitive is extremely parallelizable, each data owner would pick uniformly at random an epoch to avoid collisions. If a data owner is not writing at an epoch, they would pick a dummy database index to write to. The aggregation of all answers within a single epoch represents the final result.

\section{Related Work}

To uphold data owner privacy while performing analytics, various mechanisms have been proposed \cite{MachanavajjhalaGKV06,Sweene02,LiLV07,DBLP:conf/icalp/Dwork06,DBLP:conf/tcc/DworkMNS06}. Differential privacy has emerged as the strongest of these privacy mechanisms ~\cite{DBLP:conf/icalp/Dwork06,DBLP:conf/tcc/DworkMNS06}. The core idea of differential privacy is to provide strong bounds and guarantees on the privacy leakage when multiple aggregate analytics are run despite the presence or absence of a single data owner from the dataset. This privacy mechanism is provided by adding differentially private noise to the aggregrate answer. Differentially private systems have evolved from centralized databases, which are vulnerable to data breaches, to more modern distributed data stores. However, these modern approaches also bring various security vulnerabilities regarding the aggregation mechanism as they require trusted coordination or a centralized aggregation point. 

Recent data analytics systems achieve differential privacy guarantees in a distributed setting~\cite{DBLP:conf/eurocrypt/DworkKMMN06,DBLP:conf/ccs/HardtN12,DBLP:conf/nsdi/ChenRFG12,DBLP:conf/ccs/AkkusCHFG12,DBLP:conf/sigcomm/ChenAF13} where each data owner holds their personal data rather than in a central database. In these systems, the query answers are indepedently generated at each data owner's own device, and then the differential privacy noise is added to the aggregate answer either collaboratively by the data owners or by a centralized aggregator. However, these distributed systems require strong trust assumptions regarding the aggregation mechanism or require expensive zero-knowledge proofs to defend against pollution attacks where even a single data owner can substantially distort the aggregate result with a single malicious answer~\cite{DBLP:conf/sigmod/RastogiN10,DBLP:conf/fc/ChanSS12}.

Furthermore, these prior systems perform a one-time data collection whereby the database remains unchanged during the course of the query execution (which can stretch up to hours or days). Data analysts care about timeliness as it is common that data owners' personal data is constantly changing.  As a result, it is crucial that analysts can issue a standing query and get the timely updates continuously, i.e., \emph{stream analytics}.  These ``one-shot''  analytics systems cannot provide the required timeliness guarantees for stream analytics.  To adapt them to support stream analytics would require substantial system changes, and result in unsatisfactory loss in terms of privacy, utility, or latency.

To overcome the limitations of the aforementioned systems, several differentially private stream analytics systems have been proposed recently~\cite{DBLP:conf/stoc/DworkNPR10,DBLP:journals/tissec/ChanSS11,DBLP:conf/pet/ChanLSX12,DBLP:conf/ndss/ShiCRCS11,DBLP:conf/sigmod/RastogiN10,DBLP:conf/ndss/FriedmanSKS14,DBLP:conf/fc/ChanSS12}.  These systems inherently consider data owners' data as streams and can evolve over time.  However, these systems all have technical shortcomings, and none of them appears practical in real world.  One of the first systems~\cite{DBLP:conf/stoc/DworkNPR10} updates the query result only after data owners' data changes significantly, and does not support stream analytics over an unlimited time period.  Subsequent systems~\cite{DBLP:journals/tissec/ChanSS11,DBLP:conf/fc/ChanSS12} remove the limit on the time period, but introduce extra system overheads.  Some systems~\cite{DBLP:conf/ndss/ShiCRCS11, DBLP:conf/sigmod/RastogiN10} leverage sophisticated cryptographic operations to produce noisy aggregate query results, under the assumption that key shares have been distributed among data owners via expensive secret sharing protocols.  These protocols, however, cannot work at large scale under churn; moreover, in these systems, even a single malicious data owner can substantially distort the aggregate results without detection.  Recently, some other privacy-preserving distributed stream monitoring systems have been proposed~\cite{DBLP:conf/ndss/FriedmanSKS14,DBLP:conf/pet/ChanLSX12,DBLP:conf/ccs/ErlingssonPK14}.  However, they require some form of synchronization, and/or are tailored for heavy-hitter monitoring only (i.e., they can only report on a fraction of the data).

Various anonymity systems have been proposed all with varying trade-offs regarding throughput and trust assumptions. Chaum's Dining Cryptographers \cite{DBLP:journals/joc/Chaum88} was one of the first anonymity systems which was fully peer-to-peer and is also information-theoretically anonymous. However, the bandwidth hungry peer-to-peer coordination and the expensive pollution protection mechanisms severely limit the scalability. Recent developments such as the Dissent based systems \cite{DBLP:conf/osdi/WolinskyCFJ12,DBLP:conf/ccs/Corrigan-GibbsF10} weaken the trust assumptions slightly to improve scalability. However, while there is an improvement over the original DC-Nets, these systems are still limited by their scalability.

Chaum's mixnets \cite{DBLP:journals/cacm/Chaum81} also was an early anonymity based system that subverted traffic analysis. While mixnets only require a single trusted server, early version of mixes suffered from high-latency. Tor\cite{DBLP:conf/uss/DingledineMS04} attempted to solve the latency issue by severely introducing trust vulnerabilities. A more recent version of Chaum's mixnets \cite{DBLP:journals/iacr/DavidChaumJKKRS16} achieves low-latency though also has trust vulnerabilities. Other recent mixnets are designed to address more real-time communication \cite{DBLP:conf/sigcomm/BlondCCDM15,DBLP:conf/sosp/HooffLZZ15}. However, these systems are not strong in their cryptographic guarantees and have multiple vulnerabilities including requiring strong network assumptions such as continous cover traffic, failing to protect against timing analysis when malicious ISPs are involved, requiring the majority of parties to remain online and inability to tolerate high churn, and incurring excessive and wasteful bandwidth overheads. In general, to achieve the same cryptographic guarantees as \projecttitle, expensive zero-knowledge proofs are required which bounds the latency that can be achieved in these systems. 

Riposte \cite{DBLP:conf/sp/Corrigan-GibbsB15} is able to build anonymity sets on the order of millions by utilizing an audit server and requiring at least 2 honest servers.. This additional trust assumption greatly improves the scalability. However, the version of Riposte which is similar to \projecttitle uses a more expensive homomorphic pseudorandom generator. Riposte takes on the order of days to process compared to \projecttitle which can process on the order of seconds. An information-theoretic approach utilizing Shamir's secret sharing \cite{DBLP:journals/cacm/Shamir79} allows for the summation of multiple polynomials. The slotted approach as used in the distributed point function can also be utilized to assign a single message per data owner. However, as compared to FSS, Shamir's secret sharing requires the shares to be at least the size of the secret itself, while FSS is able to achieve a key size on the order of the square root of the key size.

\section{Conclusion}

We have presented \projecttitle, a new approach for privacy-preserving stream analytics. \projecttitle is the first such system, to the best of our knowledge, that provides all of the following in real-time: strong privacy guarantees without peer coordination or a centralized service, high accuracy, and an anonymity system resistant to traffic analysis. \projecttitle utilizes randomized response to achieve distributed differential privacy guarantees and a new cryptographic primitive named function secret sharing to enable anonymity.
\appendix

\section{Appendix}

For soundness we explain the conditional probability calculation regarding the disclosure of a sensitive attribute. Recall that our privacy mechanism utilizes  randomized response mechanism to privatize the disclosure of a sensitive attribute. A data owner has two biased coins that flip heads with probability $p$ and $q$ respectively. The data owner flips the first coin and if it comes up heads answers truthfully. If the first coin is tails, the data owner flips the second coin and replies "Yes" if heads and "No" if tails.

Let $\pi_A$ represent the probability that the data owner has the sensitive attribute. This corresponds to the underlying fraction of the population which contains the sensitive attribute.

\begin{align}
P(Yes)=p \times \pi_A +(1-p) \times q \\
P(No)=p \times (1-\pi_A)+(1-p) \times (1-q)
\end{align}

\begin{align}
P(A|Yes)=\frac{P(A) \times P(Yes|A)}{P(Yes)} \\
P(A|Yes)=\frac{\pi_A \times (p+(1-p) \times q)}{p \times \pi_A + (1-p) \times q}
\end{align}

\begin{align}
P(\neg A|Yes)=\frac{P(\neg A) \times P(Yes|\neg A)}{P(Yes)} \\
P(\neg A|Yes)=\frac{\neg\pi_A \times (1-p) \times q}{p \times \pi_A + (1-p) \times q} 
\end{align}

\bibliographystyle{abbrv}
\bibliography{fss,privacy,vehicles,mpc,main}

\end{document}